\documentclass[aps, prb, twocolumn]{revtex4-1}
\usepackage{graphicx}% Include figure files
\usepackage{amsmath}
\usepackage{amsfonts}
\usepackage{times}

\usepackage{textcomp}
\begin{document}

\title{Comparison of the density-matrix renormalization group method applied to fractional quantum Hall systems in different geometries}

\author{Zi-Xiang Hu$^1$}
\author{Z. Papi\'c$^1$}
\author{S. Johri$^1$}
\author{R. N. Bhatt$^1$}
\author{Peter Schmitteckert $^2$}
\affiliation{$^1$Department of Electrical Engineering, Princeton University, Princeton, New Jersey
08544, USA}
\affiliation{$^2$Institut f\"{u}r Nanotechnologie, Forschungszentrum Karlsruhe, D-76021 Karlsruhe, Germany}

\date{\today}
\begin{abstract}
We report a systematic study of the fractional quantum Hall effect (FQHE) using the density-matrix renormalization group (DMRG) method on two different geometries: the sphere and the cylinder.
We provide convergence benchmarks based on model Hamiltonians known to
possess exact zero-energy ground states, as well as an analysis of the number of sweeps and basis elements that need to be kept in order to achieve the desired accuracy.
The ground state energies of the Coulomb Hamiltonian at $\nu=1/3$ and $\nu=5/2$ filling are extracted and compared with the results obtained by previous DMRG implementations in the literature. A remarkably rapid convergence in the cylinder geometry is noted and suggests that this  boundary condition is particularly suited for the application of the DMRG method to the FQHE.
\end{abstract}
%\pacs{73.43.Lp, 73.43.Cd, 71.10.Pm}

\maketitle

\section{introduction}
Strongly correlated systems in low dimensions are among the most active areas in the condensed matter physics. These systems contain a large number of particles that interact strongly
with each other and cannot be understood in a single-particle picture.
A paradigm of strongly correlated systems is the fractional quantum Hall effect~\cite{tsg} (FQHE)
that occurs when a system of two-dimensional electrons partially fills one of the Landau levels
in a strong perpendicular magnetic field. Since the kinetic energy is frozen in a partially-filled Landau level, the electron-electron interaction is the only relevant term in the Hamiltonian and leads to the emergence of non-perturbative ground-states with fractionalized charge~\cite{Laughlin} and anyonic, Abelian and non-Abelian~\cite{mr}, statistics.

Due to the non-perturbative nature of the FQHE, numerical methods have played a crucial role since the original work of Laughlin~\cite{Laughlin}. In particular, exact diagonalization (ED) presented itself as a versatile and extremely powerful tool that unraveled many of the complexities of FQH systems~\cite{Haldane83, prange}. The popularity and quick success of ED was due to the specific correlations of FQH systems that rapidly minimize the finite-size effects with increasing the number of particles in the simulation. Highly accurate predictions of the system's properties in the thermodynamic limit could be obtained by considering systems as small as 10 particles~\cite{prange}.

As it is well known, the main bottleneck of ED calculations lies in the exponential explosion of the size of the many-body Hilbert space as the number of particles grows. While for the simplest FQH fractions, such as the Laughlin $\nu=1/3$ state~\cite{Laughlin}, admittedly all essential physical properties can be obtained in the systems attainable by ED, in the majority of other cases ED is not sufficient. This is particularly striking in case of spin degree of freedom, or SU(4) internal symmetry if we consider FQHE in graphene~\cite{graphene_fqhe}. However, similar constraints arise even in the spin-polarized case of the non-Abelian Read-Rezayi sequence~\cite{rr_parafermion}, where electrons are believed to pair into $k\geq 2$-body clusters. Therefore, a non-Abelian $N_e$-particle state at level $k$ is likely to have finite-size effects comparable to the Laughlin-like state of $N_e/k$ particles. 
Hence, to address the properties of the non-Abelian ground-state it is desirable to consider systems at least $k$ times as large. It is therefore of essential importance to develop new numerical methods that can reach larger system sizes than ED.

One such method is the density matrix renormalization group (DMRG), invented by White~\cite{white} in 1992. DMRG has been quite successful over the last decade when it was applied to one-dimensional systems such as the Heisenberg spin chains and the one-dimensional Hubbard model. In essence, it is a variational method to get the ground state and the low-lying energy states of the system. The algorithm contains two main parts. One is called the infinite size algorithm which grows the system to a big size, and the other one is referred to as the finite-size algorithm, which makes the ground state converge. The only approximation in the DMRG method is the truncation of the Hilbert space according to the eigenvalues of the reduced density matrix for the subblock which is
used to construct the large system. The more states are kept in the reduced density matrix, the higher the accuracy one can achieve in principle.  It is generally believed that the entanglement entropy of a subregion often grows like the boundary area of the subregion~\cite{eisert}.
A larger entanglement entropy, or larger correlation,
means that one needs to keep more states to achieve a sufficient accuracy. The success of DMRG in the one-dimensional systems was ensured by the low entanglement between the
two subregions which only have a point surface between two blocks.

On the other hand, a FQH system is two-dimensional and thus the success of the DMRG method in FQHE is by no means obvious. However, in a Landau gauge, the one-body orbitals are Gaussian-localized and provide a mapping to an effective ``one-dimensional" chain with the long-range Coulomb interaction. This motivates an attempt to apply the DMRG to the FQHE system.
The first such attempt was done for the periodic boundary conditions (torus geometry) by Shibata \emph{et al.}~\cite{shibata}, mostly considering compressible, stripe and bubble, phases in higher Landau levels.  Feiguin \emph{et al.}~\cite{feiguin} developed a DMRG scheme for the ground state and excited states in the spherical geometry for larger systems at filling factors $\nu=1/3$ and $\nu=5/2$. Hard-core interactions were also studied on thin cylinders in an unpublished work~\cite{bergholtz}, and for bosonic systems in Ref.~\onlinecite{kovrizhin}. Most recently, Zhao \emph{et al.}~\cite{Jize} developed an independent DMRG implementation that by far exceeds the previous attempts. In this study, the maximal system size was $N_e = 24$ for $\nu=1/3$ and $N_e = 34$ for $\nu=5/2$. The independent implementations~\cite{shibata,feiguin,bergholtz,kovrizhin,Jize} appear to differ significantly from each other in various aspects, in particular in the number of basis states that are kept, ranging from a few hundred in the torus geometry, up to $N_{\rm keep} = 20000$ states in Ref.~\onlinecite{Jize}.

In this paper, we report on the systematic study of the FQHE system in the spherical and cylinder geometry based on our independent implementation of the DMRG method. We address the well-studied FQH systems at fillings $\nu=1/3$ and $\nu=5/2$ with the goal of providing a detailed benchmark of the DMRG algorithm and comparing it with the previous implementations. New physical results obtained with the current DMRG implementation will be presented elsewhere~\cite{huetal}.

The remainder of this paper is organized as follows. In Sec. II we analyze the convergence of the $V_1$ Haldane pseudopotential~\cite{Haldane83,prange} Hamiltonian on the sphere that is analytically known to 
yield the Laughlin wavefunction as an exact zero-energy ground state. In the case of Coulomb interaction, we evaluate  the ground state energy for $\nu=1/3$ and $\nu=5/2$ fillings corresponding to the Laughlin~\cite{Laughlin} and Moore-Read~\cite{mr} states. The ground-state energies per particle are extrapolated to the thermodynamic limit using finite-size scaling techniques. In Sec. III, we draw some comparisons with the cylinder geometry, which is an alternative geometry for studying the FQHE that so far has scarcely been used~\cite{EDcylinder}. The convergence for the $V_1$ Hamiltonian is found to be significantly faster on the cylinder than on the sphere, suggesting that this boundary condition might be promising for further studies of the FQHE. Discussion and conclusions are given in Sec. IV.

\section{Sphere geometry}

We study a model for spin-polarized electrons moving on the surface of a sphere, with a magnetic monopole $2S$ placed in the center to generate a radially-symmetric magnetic field perpendicular to the surface~\cite{Haldane83}. In strong magnetic fields, electrons in general completely fill $(n-1)$ single particle Landau levels which are considered to be ``inert", and all dynamics comes from a partially-filled $n$th Landau level. Any two-body Hamiltonian, projected to this $n$th Landau level (neglecting the excitations to higher Landau levels), can be written in the usual second-quantized form,
\begin{equation}
 H = \frac{1}{2}\sum_{m_1,m_2, m_3, m_4} \langle m_1 m_2 | V | m_3 m_4 \rangle a_{m_1}^+a_{m_2}^+a_{m_3}a_{m_4}.
\end{equation}
In the spherical geometry, quantum numbers $m_i$'s label the $z$-component of the angular momentum for particle $i$ which takes values: $-S, -(S-1), \ldots, S$. The one-body orbitals are the monopole harmonics~\cite{Haldane83} $Y_{Slm}$ which generalize the usual spherical harmonics obtained for $S=0$.
When we target a specific many-body state, we also need to adjust the flux $2S$ to take into account the so-called shift that determines the total number of the available orbitals. This means that $2S=\frac{1}{\nu}N_e+\mathcal{S}$, where $\mathcal{S}$ is a universal number that characterizes each many-body state, e.g. $\mathcal{S}=-3$ for the Lauglin and Moore-Read state.

Because of rotational and translational invariance, any two-body interaction matrix element $\langle m_1 m_2 | V | m_3 m_4 \rangle$ can be decomposed as~\cite{fano}
\begin{eqnarray}
&&\langle m_1 m_2 | V | m_3 m_4 \rangle = \nonumber\\
&&\sum_{J=0}^{2S} \sum_{M=-J}^{J} \langle Sm_1,Sm_2|JM\rangle\langle Sm_3,Sm_4|JM\rangle V_J^{(S)}/R, \nonumber\\
\end{eqnarray}
where $V_J$ are the Haldane pseudopotentials~\cite{Haldane83} and $R=\sqrt{S}\ell_B$ is the radius of the sphere in terms of the magnetic length $\ell_B=\sqrt{\hbar/eB}$. The first two terms in the above equation are the Clebsch-Gordan coefficients on the sphere. When symmetry is taken into account, at filling $\nu=1/3$ the Lanzcos method can diagonalize the sparse Hamiltonian matrix for up to 14 electrons, corresponding to the Hilbert space dimension of $\sim 10^8$.

As shown by Haldane~\cite{prange}, the advantage of the pseudopotential formulation is that model wavefunctions can be defined as ground states of the truncated Hamiltonians. For example, the Laughlin wavefunction is obtained as an exact zero-energy ground state for the hard-core interaction with $V_1>0, V_{m>1}=0$, with an excitation gap controlled by the magnitude of $V_1$. From the computational point of view, $V_1$ Hamiltonian is nearly as sparse as the full Coulomb Hamiltonian, but it serves as a universal reference to test the accuracy of the DMRG code for large systems because the ground-state energy is known to be exactly zero for any system size.

\begin{figure}[ttt]
 \includegraphics[width=8cm]{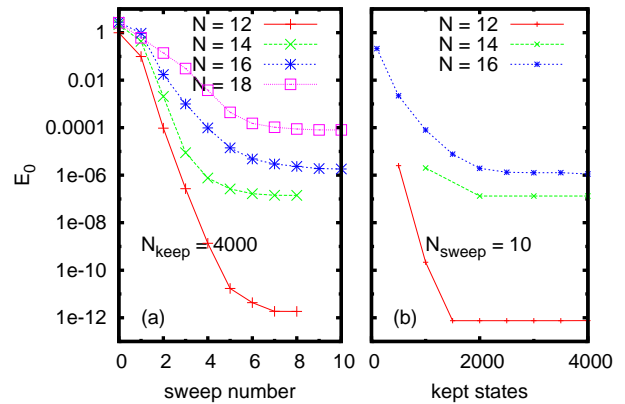}
\caption{\label{conv1}(a)The convergence of the ground-state energiesfor the hard-core $V_1$ Hamiltonian at $\nu=1/3$ as a function of the finite size sweeping number when keeping 4000 states in the subsystem. (b)The ground-state energy as a function of the number of kept states. We perform 10 finite-size sweeps for each point. The energies are on a
logarithmic scale.}
\end{figure}
In Fig.~\ref{conv1} we show the ground-state
energy convergence for different system sizes as a function of the finite-size sweep number with the fixed number of states kept ($N_{\rm keep}=4000$), or as a function of the number of kept states in the subsystem with a fixed sweep number.
For a fixed number of kept states, the accuracy of the ground state energy decreases when we increase the system size. It means that more states for the larger systems need to be kept if the same accuracy is demanded.
As shown in Fig.~\ref{conv1}(b), increasing the number of kept states obviously helps the convergence although the energy drops very slowly when $N_{\rm keep}$ is large.
However, for the largest system size with $N_e = 18$ we tested in Fig.~\ref{conv1}, the ground-state energy drops to $10^{-4}$ when just keeping 4000 states and after finishing 10 finite size sweeps.
This energy is far below the gap between the ground state and the first excited state. We assume the ground state is close enough to the Laughlin state in this case.
To improve the accuracy, one needs more finite-size sweeps and keeping more states in the truncation.

\begin{figure}[ttt]
\includegraphics[width=8cm]{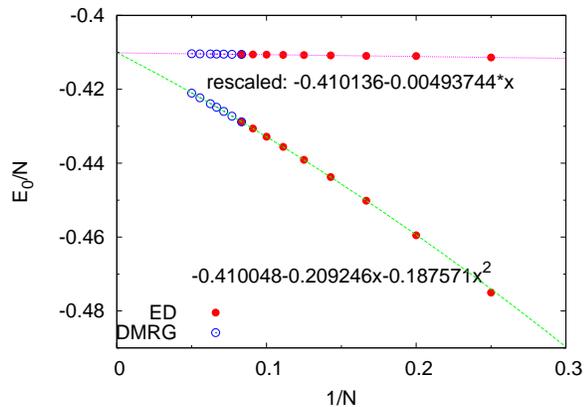}
\caption{\label{lauscaling}Ground-state energy per particle for the $\nu=1/3$ state up to 20 electrons as a function of size of the system.
Solid points are the results of exact diagonalization and the blank circles represent DMRG results.
The two fitting curves are with(upper) and without(lower) rescaling the magnetic length as described in Ref.~\onlinecite{morf1,morf2}.The energy in the thermodynamic limit is $\approx -0.4101 e^2/\ell_B$ which is consistent with previous studies~\cite{feiguin, Jize}.
}
\end{figure}

\begin{figure}[ttt]
\includegraphics[width=8cm]{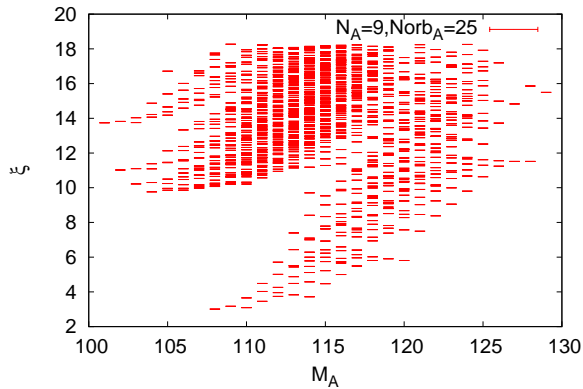}
\caption{\label{eslau}The entanglement spectrum for 18 electrons at $\nu=1/3$ with the Coulomb interaction. The subsystem contains 9 electrons in 25 orbitals.}
\end{figure}

Having established the convergence scaling for the hard-core Hamiltonian, we move to the full Coulomb interaction. At $\nu=1/3$ we calculate the ground-state energy for systems up to 20 electrons.
All the results are obtained by keeping up to 5000 states in the subsystem. With keeping the same number of states,  we find the efficiency of our code is the same as that shown
in Fig.1 of Ref.\onlinecite{Jize}. The results for different system sizes at $\nu=1/3$ are summarized in Fig.~\ref{lauscaling}, which includes the data both from the ED and the DMRG.
It shows they match with each other very well. We do the finite-size scaling for the ground-state energy per electon with a quadratic polynomial, and extrapolate the thermodynamic limit
energy to be $-0.410048 e^2/\ell_B$.
On finite spheres, it has been suggested~\cite{morf1,morf2} that the curvature effects can be substantially minized by rescaling the magnetic length $\ell_B$.
We also plot the rescaled energies in Fig.~\ref{lauscaling} and do the finite-size scaling with a linear function. The energy in the thermodynamic limit $-0.410136 e^2/\ell_B'$ is almost the same as that without rescaling the magnetic length.
This means the large-scale study by the DMRG method has already removed the finite-size effects coming from the curvature. Our results are also consistent with the previous DMRG study~\cite{feiguin, Jize}.
Besides the ground-state energy, we also plot the entanglement spectrum~\cite{lihui} in Fig.~\ref{eslau} for 18 electrons at $\nu=1/3$, for which we cut the system into two equal parts.
The splitting between the conformal part~\cite{lihui} and the non-conformal part, and the counting of the conformal states in the entanglement spectrum, demonstrate that DMRG has
captured the correct topological properties of the ground state.

\begin{figure}[ttt]
\includegraphics[width=8cm]{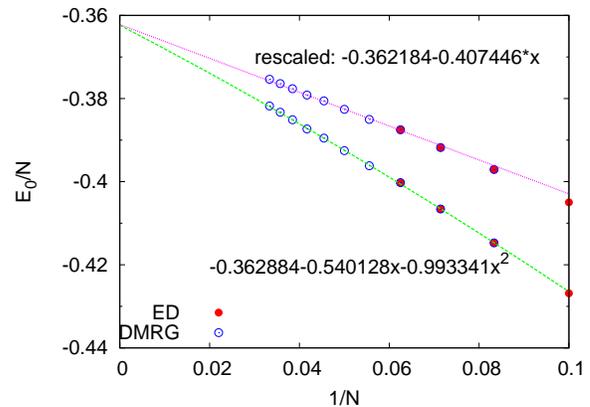}
\caption{\label{MRscal}Finite size scaling of the ground-state energy per particle for $\nu=5/2$ FQH state up to $N_e=30$ electrons.
The results are obtained by keeping up to 5000 states in the truncation and after completing  $10$ finite size sweeps.
The energy in the thermodynamic limit is consistent before and after rescaling the magnetic length $\ell_B$, which means that the large system sizes have automatically eliminated the curvature effects.}
\end{figure}
As a second case, we consider the filling $\nu=5/2$, believed to be described by the Moore-Read Pfaffian state~\cite{mr}. This state is more fragile that the Laughlin state and has a smaller gap by nearly an order of magnitude. To study the convergence, it is in principle possible to use the exact interaction that produces the Moore-Read state as a zero-energy ground state, but this is much more costly because it is a three-body interaction.
The results for the Coulomb interaction projected to $n=1$ Landau level are shown in Fig.\ref{MRscal}.
The ground-state energies are obtained for up to 30 electrons by keeping at most 5000 states. The result for the largest system size presented in this plot was obtained within
one week on a computer cluster with 12 cores and 144G memory. With the same scaling techniques as in the $\nu=1/3$ case, we extract the ground-state energy per electron in the thermodynamic limit
$\approx -0.3622 e^2/\ell_B$, consistent with Ref.~\onlinecite{feiguin, Jize}.

\section{Cylinder geometry}

\begin{figure}
 \includegraphics[width=8cm]{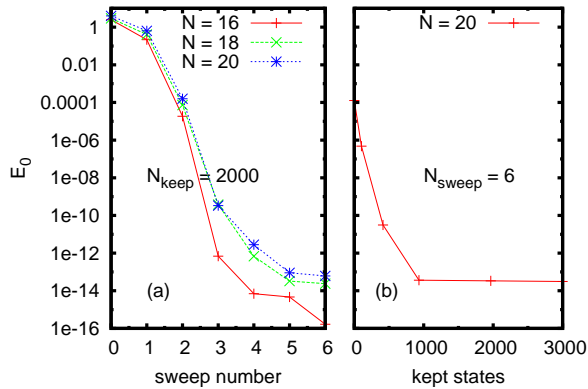}
\caption{\label{concylinder}The convergence of the ground-state energy for electrons with hard-core interaction on the cylinder. (a) We keep 2000 states in the truncation procedure for all the system sizes.
The ground-state energy drops very fast in the finite-size sweeping procedure.  (b) The dependence of the ground-state energy on the number of kept states for 20 electrons. }
\end{figure}

To complement the results obtained in the spherical geometry, in this Section we consider the cylinder geometry~\cite{EDcylinder}. Cylinder geometry is interesting because it shares  some features with the compact geometries, such as sphere or torus, but also possesses two open boundaries, which makes it convenient for the study of the edge effects, like the disk geometry~\cite{wenxg,xin02prl,xin08prb}. Compared to the sphere, the attractive feature of the cylinder is the flat surface and lack of curvature effects.

Cylinder boundary condition is compatible with the Landau gauge where periodic boundary condition in assumed along one direction (say $y$-axis) with a repeat distance $L$, and open boundary condition in the other direction ($x$-axis). The single-body wavefunction in the lowest Landau level is given by
\begin{equation}
 \psi_m(x, y) = \frac{1}{\sqrt{\pi^{1/2}L \ell_B}}e^{i k_m y} e^{-(x+k_m\ell_B)^2/2\ell_B^2},
\end{equation}
where $k_m = 2\pi m / L$ is the momentum for the $m$th orbital. The orbital index $m$ takes values $0,1,\ldots, N_{orb}-1$, and the distance in $x$-direction between two nearest orbitals is $2\pi/L$.

For a finite size system with $N_e$ particles at filling $\nu = 1/3$ for example, the
number of orbitals is $N_{orb} = 3 N_e - 2$, and thus the area of the cylinder is quantized to be $2\pi N_{orb} \ell_B^2$. To accommodate the finite number of the orbitals $N_{orb}$, we fix the extent in the $x$-direction to be $X = 2\pi N_{orb} \ell_B^2 / L$. Similar to the torus geometry, the properties of many-body states depend on the aspect ratio $\lambda = X / L =  2\pi N_{orb} \ell_B^2 / L^2$. In the following we concentrate on the FQH states that are realized in the vicinity of $\lambda = 1$.

For the $V_1$ hard-core interaction, the Hamiltonian (in units $\ell_B=1$) can be written in a simple form~\cite{EDcylinder}:
\begin{eqnarray} \label{hccylinder}
 H = \frac{1}{2}\sum_{m,n,l} (m^2-n^2) e^{-(m^2+n^2)/2} a_{n+l}^+a_{m+l}^+a_{m+n+l}a_{l}.
\end{eqnarray}
Exact diagonalization studies~\cite{EDcylinder, cylinderdelta} show that $H$ has a zero-energy ground state, but the nature of the ground state changes from the incompressible liquid to the charge-density wave upon varying the aspect ratio.
Here we focus on the liquid state and obtain it by DMRG method for large systems up to 20 electrons with the hard-core interaction.

The convergence of the ground state energy as a function of the sweep number and the number of kept states
is shown in Fig.~\ref{concylinder}.
For the system with 20 electrons, we obtain the ground state energy $10^{-13}$ when keeping only 2000 states and after completing 6 finite-size sweeps. On the other hand,
if we look at the final ground state energy as a function of the number of the kept states, we observe that the same accuracy can be reached even by
keeping only 1000 states.

\begin{figure}
 \includegraphics[width=8cm]{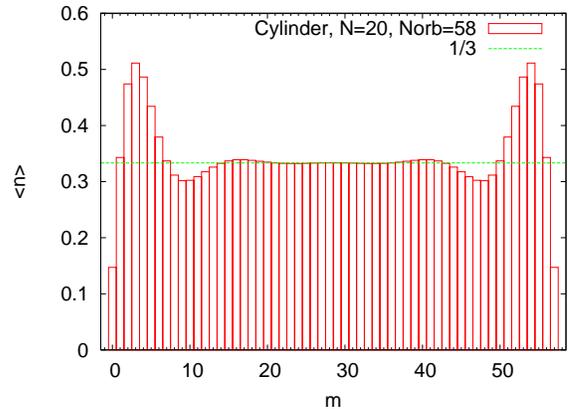}
\caption{\label{occ}The mean orbital occupation number for 20 electrons with hard-core interaction at $\nu=1/3$ on the cylinder. DMRG calculation is
performed by keeping 3000 states and finishing 6 finite size sweeps.}
\end{figure}
To verify that the ground state is indeed the Laughlin state, we plot the average occupation number $\langle c_m^+c_m\rangle$ for the system with 20 electrons in 58 orbitals in Fig.~\ref{occ}. For an incompressible liquid, the average occupation number is roughly constant in the bulk and equal to $\nu$, with some deviations close to the two edges. 
This is indeed what we observe in Fig.~\ref{occ}. 

\begin{figure}
 \includegraphics[width=8cm]{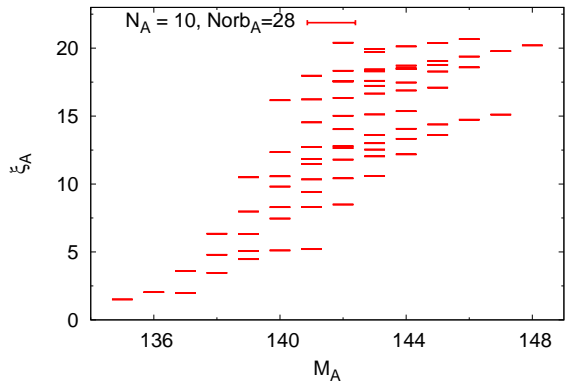}
\caption{\label{escl}The entanglement spectrum of the Laughlin state for 20 electrons on the cylinder. Because of the hard-core interaction, the entanglement spectrum only contains the conformal branch, with the same counting as in Fig.~\ref{eslau}.}
\end{figure}

We also plot the entanglement spectrum on the cylinder, Fig.~\ref{escl}. Because we use the ground-state of the $V_1$ pseudopotential Hamiltonian, the entanglement spectrum only contains a conformal branch, but is otherwise similar to the spectrum obtained on the sphere, Fig.~\ref{eslau}. In particular, the counting of the conformal levels is identical in the two cases (up to the limit set by the size of the sphere). 
Note that although the true energy spectrum reflects the presence of two edges on the cylinder, the entanglement spectrum involves only a single cut and thus probes only a single edge, in complete analogy with the sphere.

\section{Conclusions and discussion}

We presented a systematic study of the FQHE at two well-known and important filling factors, $\nu=1/3$ and $\nu=5/2$, for boundary conditions using our independent implementation of the DMRG method. In the spherical geometry, the DMRG results for the ground state energies at filling $\nu=1/3$ and $\nu=5/2$ are consistent with the exact diagonalization study for small system sizes, and the previous DMRG studies~\cite{feiguin,Jize} for large system sizes. For the largest system size we have reached, the error of the ground state energy is about $10^{-4}$ which is roughly two-three orders of magnitude below the energy gap to the excited states. The consistency in the extrapolation of the ground-state energy shows that these system sizes have negligible curvature effects.

The application of the DMRG method to the cylinder geometry shows much higher efficiency compared to the sphere. Based on the convergence for the $V_1$ interaction, we expect the cylinder to be the more promising venue for the future applications of DMRG. Due to the presence of two open edges, the treatment of the full Coulomb interaction is not as straightforward as in the compact spherical geometry, and requires special care in defining the confining potential to contain the fluid. One may furthermore expect various phase transitions as a function of the aspect ratio and the magnitude of the confining potential relative to $e^2\epsilon\ell_B$. Details of these studies will be presented elsewhere~\cite{huetal}.

\section{Acknowledgments}
We would like to thank E. H. Rezayi, F. D. M. Haldane for simulating discussions. Z-X. Hu also thanks Jize Zhao for comparing the results on the sphere.
This work is supported by DOE grant No. DE-SC0002140.

\end{document}